\documentclass{aastex}     %% The manuscript based on AASTeX v5.x
\usepackage{spr-astr-addons}  %% mimicing ASTR journal style
\usepackage{url}
\usepackage{color}
\usepackage{comment}
\usepackage{epstopdf}
\usepackage{gensymb}

\begin{document}
%% Article title
%
%\title{<Article title>}
\title{Analysis of the exoplanet containing system Kepler-91}
%% Running heads

\shorttitle{Analysis of Kepler-91}
\shortauthors{<Budding et al.>}

%% Author and Affilations
\author{E.\ Budding\altaffilmark{1,2,3,4}}
\and
\author{\c{C}.\  P\"{u}sk\"{u}ll\"{u}\altaffilmark{1}}
\and
\author{M.~D.\ Rhodes\altaffilmark{5}}
\and
\author{O.\ Demircan\altaffilmark{1}} 
\and 
\author{A.\ Erdem\altaffilmark{1}}

%\affil{ University of Canakkale, TR 17020, Turkey;}
%\email{ed.budding@gmail.com} %% non-output
%\author{M.~D.\ Rhodes\altaffilmark{1}}
%\affil{BYU, Provo, Utah} 
%\and 
%\author{E.\ Budding\altaffilmark{2,3,4,5}}
%\affil{ University of Canakkale, TR 17020, Turkey;}
%\email{ed.budding@gmail.com} %% non-output

%% Alternate Affilations

\altaffiltext{1}{University of Canakkale, TR 17020, Turkey;}
\altaffiltext{2}{Carter Observatory and}
\altaffiltext{3}{SCPS, Victoria University of Wellington and}
\altaffiltext{4}{Dept.\ Physics \& Astronomy, UoC, New Zealand.}
\altaffiltext{5}{BYU, Provo, Utah}
 
\vspace{2mm} 
%EndAName
%$^{1}$ University of Canakkale, TR 17020, Turkey;\\
%$^{2}$ Department of Physics and Astronomy, University of Canterbury;\\
%$^{3}$ SCPS, Victoria University of Wellington; \\ 
%$^{4}$ Carter Observatory; New Zealand; and \\
%$^{5}$ Brigham Young University, Provo, Utah, U.S.A.
%}
%% Abstract
\begin{abstract}
We have applied the graphical user interfaced close binary system analysis program  {\sc WinFitter}  to
 an intensive study of Kepler-91 using all the available photometry
 from the NASA Exoplanet Archive (NEA) at the Caltech website: 
 
 {\footnotesize
{\bf \noindent http://exoplanetarchive.ipac.caltech.edu.  }}

 \noindent Our fitting function for the tidal distortion derives from the relevant Radau equation
 and includes terms up to the fifth power of the fractional radius.
 This results in a systematic improvement in the mass ratio estimation over
 that of Lillo-Box et al.\ (2014a) and our derived value for the
 mass ratio is in close agreement with that 
 inferred from recent high-resolution spectroscopic data.
 
 It is clear that the data analysis in terms of simply an eclipsing binary system
 is compromised by the presence of significant other causes of light variation,
 in particular non-radial pulsations. We apply a low-frequency filtering
 procedure to separate out some of this additional light variation.
 Whilst the derived eccentricity appears then  reduced, an eccentric
 effect remains in the light curve.  We consider how this
 may be maintained in spite of likely frictional effects operating over a long
 time. There are also indications that could be associated with Trojan
 or other period-resonant mass concentrations.
 Suggestions of a possible secular period variation are 
 briefly discussed.
 
\end{abstract}

%% Keywords

%% Please use labels (\label, \ref) for section, figure, table, 
%% equation reference. Use \cite for bibliography references.
%
%\section{}%\label{s:?}
%\subsection{}%\label{ss:?}
%\subsubsection{}%\label{sss:?}

\section{Introduction}

Astrophysical aims of the {\em Kepler} Mission, together with practical details
on its construction were set out by Borucki et al.\ (2003).  
Devore et al.\ (2009) provided general motivational background,
noting the 400 year anniversary of Kepler's publication of the laws of
elliptical orbits and equal areas in the year of the satellite's launch.
The Ames Research Center has had a prominent role in the materialization of such purposes. A comprehensive early summary was that of Borucki et al.\ (2011).
In mid-2013 NASA announced that two out of the original four reaction wheels used in pointing the telescope on the satellite had become inoperable and the initial objectives of the mission would be compromised as a result. However, there is, by now, a large archive of photometric data inviting continued close attention and discussion.

Rhodes \& Budding (2014), giving further background information relevant also to the present paper, tested their light-curve fitting software for {\em Kepler} exoplanet transit light curves against results published by others.  They  concluded that there was a fair measure of agreement about published parameter values although significant differences were found in some cases.
Many of these differences can be associated with the transition 
of the photometric to inferred absolute parameters.  Rhodes \& Budding (2014) accounted for this in terms of the high sensitivity of 
derived mean densities to imprecisely known, but observationally obtained,
surface gravities.   In the case of KOI 13.01, however, it is feasible that
the differences in derived inclinations {\bf are valid and can be 
associated with real short term changes due to precession} (Szab\'{o} et al., 2013). KOI 377.01 (= Kepler-9) is similarly open to 
possible variations of empirical parameters associated with orbit complexities of this known multi-planet system. For KOI 3.01, the difference in the derived stellar relative radius ($r_1$) may again arise
from orbital complexity (Bakos et al.\ 2010).  The unusually high
limb-darkening coefficient found for Kepler-1 has remained with
further study of the data and is an issue calling for continued
attention and analysis, perhaps related to other peculiarities
associated with this hot jupiter (Kipping \& Spiegel, 2011).

 The fitting program {\sc WinKepler}, discussed in Rhodes \& Budding (2014), has since been upgraded and is now called {\sc WinFitter}. It performs optimization by a modified Marquardt-Levenberg application of a fitting function to a photometric data-set (light curve). The fitting function is based on the {\em Radau model} developed from Kopal's (1959) approach to the tidal and rotational distortions (ellipticity), together with the radiative interactions (reflection), of massive and relatively close gravitating bodies. It can be downloaded from {\bf {\footnotesize http://home.comcast.net/$\sim$michael.rhodes/. } }

 The upgrade allows the user the option of regular stellar eclipsing binary light curve analysis as well as that of exoplanet light curves such as those released by the {\em Kepler} Mission via the Mikulski Archive for Space Telescopes (MAST). 

Optimal models produced by {\sc WinFitter} correspond to the least value of $\chi^2$, defined as $\Sigma(l_{o,i} - l_{c,i})^2/\Delta l_i ^2$ (Bevington, 1969), where $l_{o,i}$ and $l_{c,i}$ are the observed and calculated light levels at a particular phase. $\Delta l_i$ is an error estimate for the measured values of $l_{o,i}$. The NASA Exoplanet Archive
(NEA) lists empirical values of $\Delta l_i$ for each
datum, which guide the values assigned
to {\sc WinFitter}.  Given such starting $\Delta l_i$s,
we have retained flexibility in the
values to be used in the $\chi^2$ calculation, however, 
as it became clear that there
are sources of photometric variation other than the eclipsing binary
effects and photon noise. The question of what can be regarded as
random noise, from the point of view of systematic effects on the close eclipsing system model, is a matter we discuss further in Section 2.2. 
 
 The theoretical light-level $l_c$ corresponds to the given fitting function. The central 
 problem in the analysis of photometric data, like that discussed in the present  paper, is to find the best values of the parameters of this fitting function, together with the demonstration of a formally determinate optimal set of such parameters, whilst ensuring adequacy of the underling model to account for the observed effects. Location of a single optimum in the hyperspace formed by the unknown parameters and $\chi^2$ is effected by the simultaneous vanishing of the $\chi^2$ derivatives with respect to all those sought parameters. Determinacy is checked by the $\chi^2$ Hessian staying positive definite in the vicinity of this optimum. Adequacy of the model entails that the value of $\chi^2/\nu$, where $\nu$ is the number of degrees of freedom of the data-set, should be acceptably close to unity at the optimum. Standard tabulations of the $\chi^2(\nu)$ values at given probability levels can be used to check this. Rhodes \& Budding (2014) included further discussion of this subject as well as the physical basis of the modelling (cf.\ also Budding \& Demircan, 2007 (Ch.\ 9); Rhodes, 2015).

\section{Kepler-91}
 
\noindent {\bf 2.1 \hspace{2em} Preliminary information}

Kepler-91 (= KOI 2133.01; KIC 8219268
\footnote{KIC stands for Kepler Input Catalogue, KIC for Kepler Object of Interest.
The webpage 
http://exoplanetarchive.
ipac.caltech.edu/docs/data.html?redirected
guides users to numerous data-files, including the `cumulative list'
that we used to obtain prior parameter estimates.
The official website for {\em Kepler} light curves is actually
https://archive.stsci.edu/kepler/downloads{\_}options.html.
However, the NASA Exoplanet site referred to above does not require
file-type conversion, and has convenient normalization and 
useful plotting options. The same light curve data can be downloaded from either site.}
is believed to contain an unusual combination of a post-Main-Sequence star ascending the Red Giant Branch  accompanied by 
a jupiter-sized planet heated to around 2000 K
 (Huber et al., 2013; Lillo-Box et al., 2014a). Planetary transits are not immediately apparent in individual long-cadence data-sets: the identification 
was achieved by Tenenbaum et al.\ (2013) after applying special periodicity-finding procedures.
We used the ephemeris attributed to Tenenbaum et al.\ (2013), as
listed in the NEA, in phasing
the data and the adopted epoch and period are given in Table 1.  {\bf In doing this
we have assumed that the
adopted Epoch does indeed correspond to an identified transit mid-minimum.}
 Some tens of giant planet + giant star configurations have been found hitherto and they have attracted interest related to the theory of their origin and evolution (cf.\ e.g.\ Lin et al., 1996; Johnson et al., 2007; Nagasawa et al., 2008; Villaver \& Livio, 2009). 

An aspect of this subject concerns possible engulfment of a close planet as the host star expands at the end of its Main Sequence stage. That Kepler-91b\footnote{The symbol b is associated with the hot jupiter component. 
Kepler-91 may refer to the entire system or sometimes just the host star.} is a hot jupiter in such a situation was confirmed by Lillo-Box et al.\ (2014a,b), and more recently by Barclay et al.\ (2015), though that picture had been contested by Esteves et al.\ (2013), and subsequently by Sliski \& Kipping (2014).  More recently, Esteves et al.\ (2015) have analysed a larger data-set and retracted from their previous findings.  Their latest results are included in Table~4 for comparison.
Some of the earlier discussion related to 
density estimates, whose sensitivity to error may be associated with the 
cubed parameter $r_1$ in the denominator of
the constraint $\rho_\star = 3/4\pi G P^2 {r_1}^3$.  If an
absolute radius $R_\star$ is estimated from separate evidence, the density
is alternatively constrained by $\rho_\star = 3g/4\pi G R_\star$, but again a relatively large proportional error in the observationally determined
value of the surface gravity $g$ would compromise the density estimate
 (Muirhead et al., 2012; Rhodes \& Budding, 2014).  

 Lillo-Box et al.\ (2014a), using the 2.2 m telescope at Calar Alto (Spain), checked for the possibility of background contamination associated with the relatively large angular size of the {\em Kepler} pixels ($\sim$4 arcsec). They also looked for possible non-planet-related trends in individual data segments.  Commonly used flux measures have been tabulated by the
 archive data-source under the heading PDCSAP (pre-search data conditioning simple aperture photometry) fluxes.  These data have resulted 
 from additional processing after the SAP data (also tabulated), 
 with the aim of mitigating data artefacts.  Sometimes, 
 investigators apply separate correction procedures on the SAP data,
 but Lillo-Box et al.\ (2014a) deemed it sufficient to utilize
 the listed PDCSAP information.
   Barclay et al.\ (2015) also started with the PDCSAP fluxes, though they subsequently modelled and separated out systematic effects in the data. 

In preparing their analysis of the system, Lillo-Box et al.\ (2014a) noted the importance of reliable host star parameters, particularly where these can be elucidated by separate means prior to study of the {\em Kepler} photometry. They made use of high resolution spectroscopy of Kepler-91 from Calar Alto to this end, fitting, for example, model spectral energy distributions, matching individual spectral lines and including detailed asteroseismological analysis. In this way, they were able to give estimates of the radius and mass values as 6.3$\pm$0.16 R$_{\odot}$ and 1.31$\pm$0.10 M$_{\odot}$. Other relevant physical attributes were given too, including an age, estimated to be not far from 5 Gy, though with a substantial margin of error. Table 2 of Lillo-Box et al.\ (2014a) collects together stellar parameters from a variety of previous sources, where a fairly wide range of values can be noted (appreciably larger than the quoted errors). Our Table 1 assembles input information needed to build up a more comprehensive picture of the system.

The fact of pulsational activity in this star, relating to the aforementioned asteroseismological work, could be anticipated on the basis of the arguments used in Rhodes \& Budding (2014) concerning the level of scatter observed around the mean value of the stellar flux. Rhodes \& Budding (2014) found that for a steady source affected only by Poissonian noise, this should work out at $\sim\psi/23.8$, where the Poisson factor $\psi = 1/\sqrt{N_f}$, $N_f$ being the mean PDCSAP flux count, in this case for the long cadence data-sets. This would lead to a datum probable error of about 0.00012 (120 ppm), whereas what is found in typical light curve fitting is a scatter (regarded as noise) of about
3 times this level. There are thus clear indications in these data-sets of effects
that regular eclipsing binary modelling does not take into account. The analysis of Barclay et al.\ (2015) addressed this point in some detail, though their light-curve rectification to an eclipsing binary model was on an essentially empirical basis. 

\vspace{2ex} 
\noindent {\bf 2.2 \hspace{2em} {\em Kepler} photometry analysis}

\noindent {\em 2.2.1 Preparation}

In our analysis of this system using  {\sc WinFitter}, we used all available long cadence data sets for the observing runs (quarters) 1-17, which we downloaded from the NEA. For each data-set we processed the listed data points using the period of Tenenbaum et al.\ (2013) and normalized the PDCSAP fluxes to unity. The given times of observation (BKJD) were then converted to phases in the range 0.0 to 1.0. In this way, we constructed working light curves of $\sim$4000 points.
These data-sets proved informative, although they appear very scattered to the eye.

For initial values of the input parameters, we could use information from the KOI (Kepler Object of Interest: cumulative list) given in the NEA (see also Batalha et al., 2013).
We also take into account the parameters listed by Lillo-Box et al.\ (2014a), whose emphasis on prior host-star data was mentioned above (see also Seager \& Mall\'{e}n-Ornelas, 2003). Our Table 1 numbers
are thus rounded averages from these sources.
A more detailed discussion about prior possibilities for such input information was given by Barclay et al.\ (2015), whose analysis procedure allows for variation of hyperparameters.
However, the main parameters connecting with photometric effects, such as the eclipse and its shape, constitute separate information to absolute parameter specification. Fittings to the transit minima led to reasonably consistent indications about $r_1$, $k$ and $i$ (the radius of the star expressed as a fraction of the semi-major axis of the orbit, the ratio of planet to stellar radii, and the orbital inclination) independently of much of the information set out in Table 1.  The inclination turns out to be relatively low ($\sim$70\degr\ in Table 2) and the star is a relatively large fraction ($\sim$0.4, Table 2) of the planet-star separation, so that a noticeable boundary correction (cf.\ Kopal, 1959; ch.\ 4) can occur for moderate levels of distortion.  This can delay the onset of the transit by a degree or two of phase from what would correspond to a circular stellar disk. The oft-cited formulae of Mandel and Agol (2002), used in transit analyses such as those mentioned above, could thus render the fitting function less than adequate for precise data, regardless of the precision of available priors. 

Our general procedure is to approach the optimal photometric parameter specification in steps. First we concentrate on the parameters that have a relatively strong effect or are highly constrained by the light curve's form. When these settle towards well-defined numbers, we allow for simultaneous improvements in weaker parameters. In preliminary fittings we would generally start by  optimizing the scaling constant for the light curve's vertical axis (unit of light) $U$,\footnote{In principle, the adopted value of $U$ could be used to scale the representative flux from the system and
thence derive a corresponding stellar magnitude.  However, probably related to the wide passband of
the {\em Kepler} filter, the calibration in question turns out not simple, and we can see an appreciable
departure from the trend of magnitudes with effective wavelength listed in Table 3 of Lillo-Box et al (2014a) in the
case of the {\em Kepler} magnitude. Our flux-derived {\em Kepler} magnitude (12.408) thus appears too faint at the flux-weighted wavelength 0.735 $\mu$m, as judged by the run of values cited by Lillo-Box et al.\ (2014a).} and similarly the $x$-axis zero point (epoch of eclipse mid-minimum), or correction
to the assigned phases $\Delta\phi_0$. We then use such values in subsequent runs, keeping the generally independent $\Delta\phi_0$ fixed, and optimizing for $U$, $r_1$, $k$; and $i$. 
 
In Figure 1 we show our optimal fitting to the transit region.  We extracted the transit portions of the light curves corresponding to the phase 
range --0.05 to 0.05. Figure 1 shows the result of combining all the transit regions of all 17 light curves into a single sorted data-set of 6322 points, which has been then binned to a representative 97 points in bins of phase
interval 0.001.  
Lillo-Box et al.\ (2014a) report a procedure of initial transit analysis: their Fig 6 may be compared with our Fig 1.

\begin{figure}[H]
\begin{center}
\includegraphics[width=\columnwidth]{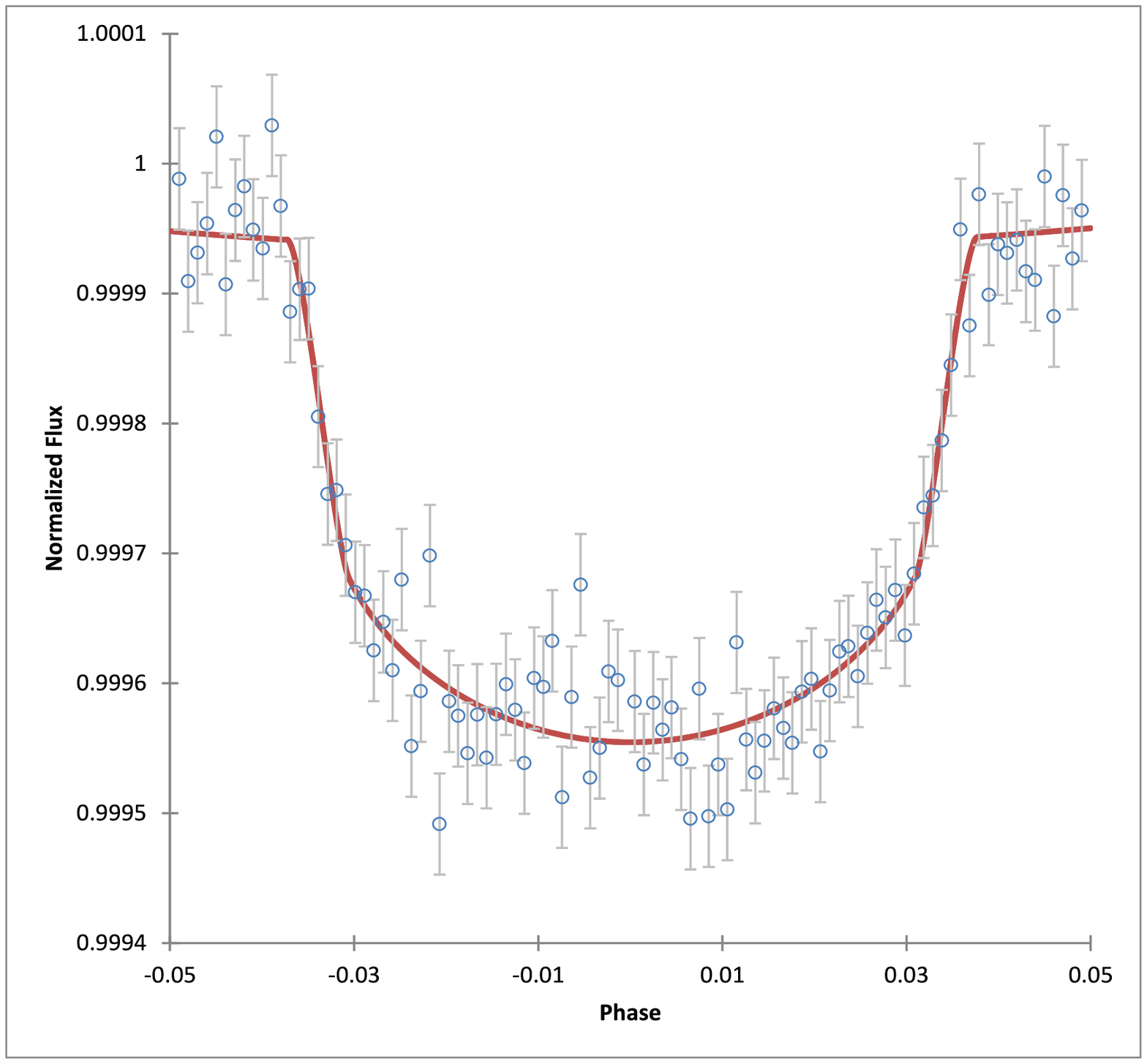} 
\label{fig:Kep91transit}
\end{center}
\caption{Binned data covering the transit region for Kepler-91b, matched by the {\sc WinFitter} model.
This may be compared with Fig.\ 6 in Lillo-Box et al.\ (2014a),
or Fig.\ 4 in Esteves et al.\ (2015).}
\end{figure}

\vspace{2ex}
\noindent {\em 2.2.2 More detailed examination}

Using the averaged values for the main geometric quantities from the initial fittings (bottom row of Table 2), 
we proceeded to examine the complete light curves. It can be readily expected from the preliminary 
information that proximity effects due to the relatively low separation of this hot 
jupiter should be detectable, although the system is fairly faint ($V$ = 12.884, Everett et al., 2012) 
and the scatter of the data points relatively high, as noted above. 
In subsequent fittings to complete light curves we experimented by concentrating all the 
determinability of the data to optimizing only $U$ and the mass-ratio, $q$, the other parameters 
having been fixed from previous fittings. This procedure is sometimes called a photometric $q$-search.  The essential form of the fitting function 
used here
is as given in Eqn (9.17) in Budding \& Demircan (2007), with further details being given in the bibliographical notes (9.7) of that book. {\sc WinFitter}  
has now multiplied the relative luminosities {\cal L}$_{1,2}$ of the original formulation by `Doppler-beaming' factors, as explained in 
the paper of Shporer et al (2012).  

We continued with further fittings for the 17 light-curves folded over each quarter, allowing the geometric 
parameters to relax from previously found values. Table 2 summarizes the findings of a large number
of fitting experiments as performed by our team-members separately.
 The run of values through the tables allows an insight into the determinability of the parameters, 
 apart from the formal errors of each fitting. The 17th data set contains only about 1/4 of typical 
 quarters and its best-fitting values diverged somewhat from the trend of the other solutions, while the reduced $\chi^2$ remained relatively high. We have therefore included the results from that fitting in the tabulated mean values
 with a correspondingly reduced weight.   At this point, we had failed to confirm any eccentricity 
to the orbit from these separate raw data-sets through finding a  significant improvement in $\chi^2$ 
from inclusion of optimized $e$ and $\omega$ values.

\begin{figure}[H]
\begin{center}
\includegraphics[width=\columnwidth]{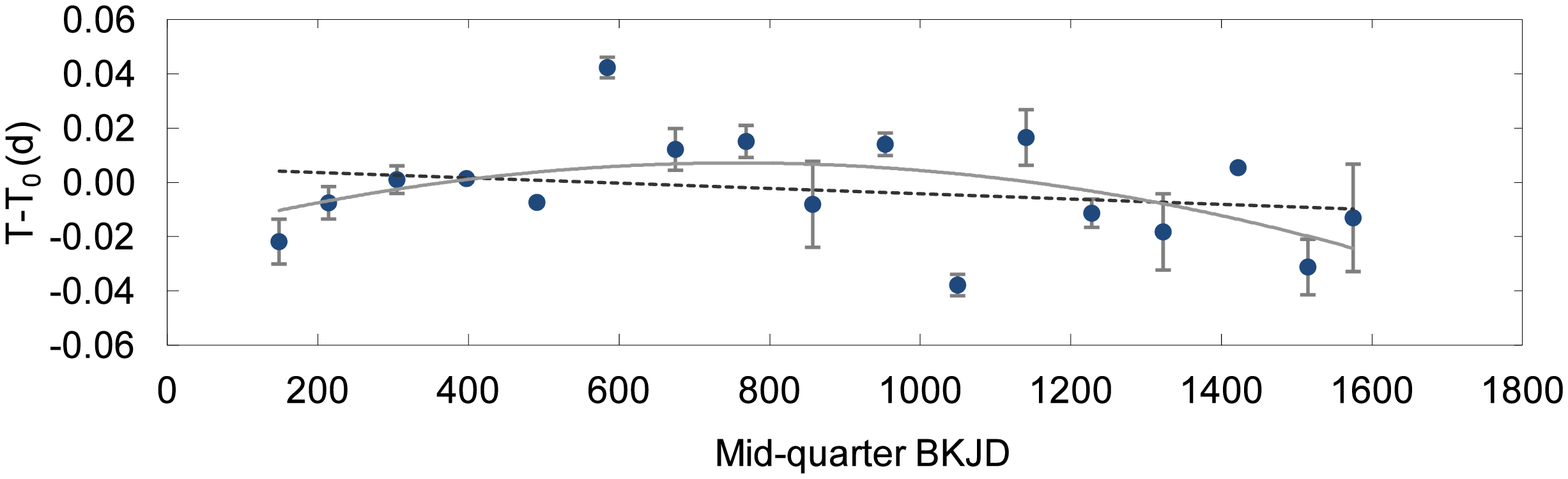} 
\label{fig:delphiomc2}
\end{center}
\caption{Differences in observed and determined times of minimum for 
Kepler-91.
The straight line (dashed) fit is suggestive that the assigned period may have
been too long -- by about 0.00001 d -- but the resulting improvement is not significant.
Somewhat greater significance (in a $\chi^2/\nu$ sense) attaches to the secular period
decrease (parabola), but even there the improvement in probability is not high (see text).}
\end{figure}

We also looked for any trend that might be discernible in the values of $\Delta \phi_0$, which
relates to the time of transit variation (TTV) exhibited in some exoplanet systems.
Our results on this are indicated in Figure 2 with the corresponding numbers in column 3 of Table 2\footnote{It should be noted that 
the effects shown in Fig.\ 2 are not
timing variations of individual transits, but an average result for each
quarter. Any trend that might be discerned would not be commensurate 
with the orbital period therefore.}.
The inclusion of a linear correlation (incorrectly assigned period) improves the value
of $\chi^2$, but only marginally, and, in fact, $\chi^2/\nu$ is not significantly altered.
A more noticeable reduction in $\chi^2/\nu$ (19\%) 
is obtained by the parabolic fit (decreasing period)
but the decrease is from around a one in three ratio of exceeding the higher $\chi^2/\nu$ by pure chance,
to around a one in three ratio of being less than the lower $\chi^2/\nu$ value randomly, and so
cannot be considered significant.  Alternatively, the Bayesian information
criterion (BIC), which is easily related to the $\chi^2$ variate
using the formula of Kass \& Raftery (1995), gives a very small
preference for simply keeping the adopted period with a slight change
of the reference epoch.  But the first 3 fittings
(reference point only, linear, and parabolic trends, have 
no significant difference in corresponding BIC values (23.3, 25.0, 24.2;
 respectively), though a 4-parameter fit  (BIC = 27.0) 
 would not be supported, by the assessment table of Kass \& Raftery.
 
In fact, the implications
of such a period reduction would be the highly unexpected circumstance of catching a planet  within
a few thousand of years before its demise. On the contrary, Sato et al (2015) gave a slightly increased period
value from that given by Tenenbaum et al.\ (2013) (see Table 4).  There are, however, interesting alternative physical scenarios
relating to period changes that we discuss later.

\begin{table*}[]
{\footnotesize

\begin{center}
%\begin{Large}
\hspace{2em} \caption{Primary Input Data \label{tbl-1}}
\begin{tabular}{|r|r|r|r|r|r|r|r|r|r|r|r|}
\hline 
\multicolumn{1}{|c|}{KOI} & 
\multicolumn{1}{|c|}{$V$} &
\multicolumn{1}{|c|}{$M_*$} & 
\multicolumn{1}{|c|}{$R_*$} & 
\multicolumn{1}{|c|}{$T_*$} & 
\multicolumn{1}{|c|}{Epoch} & 
\multicolumn{1}{|c|}{$P$} & 
\multicolumn{1}{|c|}{$Z$} & 
\multicolumn{1}{|c|}{$\log g$} & 
\multicolumn{1}{|c|}{$a$} & 
\multicolumn{1}{|c|}{ ${M_p}/{M_*}$} & 
\multicolumn{1}{|c|}{$u$}  \\ 
\hline 
2133.01 & 12.884 & 1.3 & 6.4 & 4600 &2454969.3966& 6.246580 & 0.11 & 2.94 & 0.073 & 0.00005 & 0.74 \\ 
\hline 
\end{tabular} 
\end{center}}
\vspace{1ex}
\begin{footnotesize}
\noindent
 We have here used the notation: 
$V$ -- conventional $V$ magnitude,
$M_*$ -- mass of star (solar masses),
$R_*$ -- radius of star (solar radii),
$T_*$ -- temperature of star (K),
$P$ –- orbital period (in days),
$Z$ -- metallicity of star,
$\log g$ -- log$_{10}$ of the surface gravity of star,
$a$ -- semi-major axis in AUs,
$M_p/M_*$ -- ratio of planet to star masses, 
$u$ -- stellar (linear) limb-darkening coefficient.
The numbers given in this table are rounded averages
from the results of Lillo-Box (2014a) and various sources
provided by the NEA, including the cumulative list, 
the confirmed planets and exoplanet transit survey data.

\end{footnotesize}
\end{table*}

\begin{table*}
\begin{center}
\caption{Results of optimal fittings to light curves drawn from the 17 available quarters.
\label{tbl-2}} 
{\footnotesize 
\begin{tabular}{lrrrrrrr} 
 \multicolumn{1}{l}{Qtr}& \multicolumn{1}{c}{$U$} & \multicolumn{1}{c}{$\Delta \phi_0$} 
 & \multicolumn{1}{c}{$r_1$} & \multicolumn{1}{c}{$k$} 
 & \multicolumn{1}{c}{$r_2$} &	\multicolumn{1}{c}{$i$(deg)}&	
 \multicolumn{1}{c}{$q$} \\
 \hline \\	 
1 &1.00079 & --1.5 & 0.404 & 0.0261 & 0.0105 & 68.2 & 0.00039 	 \\
2 &1.00084 & --0.5 & 0.406 & 0.0247 & 0.0100 & 68.5 & 0.00040  \\	
3 &0.99972 &  0.2 & 0.403 & 0.0209 & 0.0084 & 69.1 & 0.00037  \\	
4 &0.99987 &  0.3 & 0.403 & 0.0206 & 0.0083 & 70.0 & 0.00021  \\
5 &0.99985 & --0.6 & 0.403 & 0.0229 & 0.0092 & 70.1 & 0.00039  \\		
6 &0.99991 &  2.0 & 0.407 & 0.0228 & 0.0092 & 68.6 & 0.00065	 \\
7 &0.99980 &  0.6 & 0.403 & 0.0183 & 0.0074 & 69.8 & 0.00048	 \\
8 &0.99981 &  0.6 & 0.400 & 0.0224 & 0.0089 & 70.4 & 0.00046	 \\
9 &0.99984 & --0.4 & 0.393 & 0.0215 & 0.0084 & 70.3 & 0.00027  \\		
10 &0.99984&  0.7 & 0.406 & 0.0229 & 0.0093 & 69.1 & 0.00051  \\	
11 &0.99988& --2.1 & 0.405 & 0.0237 & 0.0096 & 68.9 & 0.00054  \\		
12 &0.99984&  0.5 & 0.396 & 0.0220 & 0.0087 & 69.5 & 0.00042  \\
13 &1.00000& --0.6 & 0.395 & 0.0220 & 0.0087 & 70.1 & 0.00012  \\
14 &1.00002& --0.9 & 0.403 & 0.0221 & 0.0089 & 69.6 & 0.00036  \\
15 &0.99994&  0.6 & 0.405 & 0.0206 & 0.0084 & 68.8 & 0.00056  \\
16 &0.99967& --1.7 & 0.407 & 0.0213 & 0.0086 & 69.9 & 0.00048  \\
17 &0.99948& --0.5 & 0.409 & 0.0196 & 0.0080 & 69.2 & 0.00055  \\
  &    &    &    &    &    &   &    \\	
Mean &0.99999(34)& --0.2(1.0) & 0.4026(43)& 0.0220(18) & 0.0089(7) & 69.4(7)& 0.00045(6) \\	
\hline
\end{tabular}
}
\end{center}
\end{table*}
 \vspace{2ex}

 We next combined all $\sim$65000 data points into a representative binned light curve that we fitted
in various ways, starting by using the mean parameters given at the bottom of Table 2.  
The results are summarized in Table 3. 
%and a representative fitting is shown in Figure 3. 
 We can note here the good measure of agreement on the main geometric
 parameters ($r_1$, $k$ and $i$), as well as the mass ratio, from 
 the various fitting approaches.
   
Lillo-Box et al.\ (2014a) carried out a basically similar procedure, aiming also to determine a planet to star mass
 ratio, and thence, using their separate information on the star, a planetary mass. Their fitting function combines the three contributions of ellipticity, Doppler beaming (a relatively very small term) and reflection, given by their Eqns 7, 8 \& 9, respectively. Eqns 7 and 9 are very simple, however,and
  although their key ellipticity contribution derives originally from the same
Radau-model formulae that we use (Kopal, 1959), the neglect of the third and fourth order terms in the tidal distortion (given the large relative size of the star) turns out to lead to a significant overestimate of the mass ratio as determined from the photometric effect of the stellar tide.  Given that $r_1 \approx 0.4$ we can see that the neglect of the third and
fourth order terms will bring their mass ratio (0.00064) down to a value in close agreement with ours.
%Lillo-Box et al.'s value for relative stellar radius appears somewhat high in comparison to ours,
% which is probably also related  to this curtailing of the higher power terms in $r_1$.

 Lillo-Box et al.\ (2014a) gave the planet's mass to be about 0.88  +0.17/--0.33 M$_{\rm jup}$ but with radius about 1.384 +0.11/--0.054R$_{\rm jup}$, i.e.\ a density of about 1/3 that of Jupiter. This was interpreted as an atmospheric inflation, associated with the relatively high stellar irradiation. They also reported an eccentric orbit ($e$ = 0.066 +0.013/--0.017, $\omega$ = 316.8+21/--7.4 deg), associated primarily with an apparent asymmetry of the light curve as a whole. They could not confirm this definitely from asymmetry of the transit alone, accepting that the eclipse effect was inconclusive regarding eccentricity. Our findings
 are similar on this point (Table 3).
 The complete light curve they modelled consists of about 260 individual points that have been binned from the original 
 complete data-set, from which they had clipped outliers so as to arrive at a working sample of $\sim$52000 points. From our fittings to the original data-sets we estimated the s.d.\ scatter of such data at $\sim$0.00037, hence the error-bars shown on the light curve of Lillo-Box et al.\ (2014a) should be typically around 0.00002 on a Poissonian assumption of their distribution. This corresponds to the error bars shown in their Fig 7.

It is clear from that Fig 7, however, that the residuals are not distributed randomly around the model curve with this precision. The model curve goes through quite fewer than $\sim$2/3 of the shown error bars, that one could reasonably expect it would do on the basis that the data corresponds only to the close binary model plus random noise. But this would happen anyway as a result of the pulsational effects. This matter was addressed by Barclay et al.\ (2015), who introduced an additional empirical noise model that they attributed to the effects of surface granulation of the host star. We digitized the data points of Fig 7 in Lillo-Box et al.\ (2014a) and could mostly confirm their model parameters to within reasonable errors including a similar eccentricity result.
In order to probe the situation further we  binned the source data and applied {\sc WinFitter} to the resulting light curve. 

%Figure 3 shows our binned light 
%curve and adopted model. Our parameters are listed in Table 3, where comparisons are made with other results.
%The fittings to the complete light curve are clearly improved by admitting a small eccentricity to the photometric effects.

\begin{figure}[H]
\begin{center}
\includegraphics[width=\columnwidth]{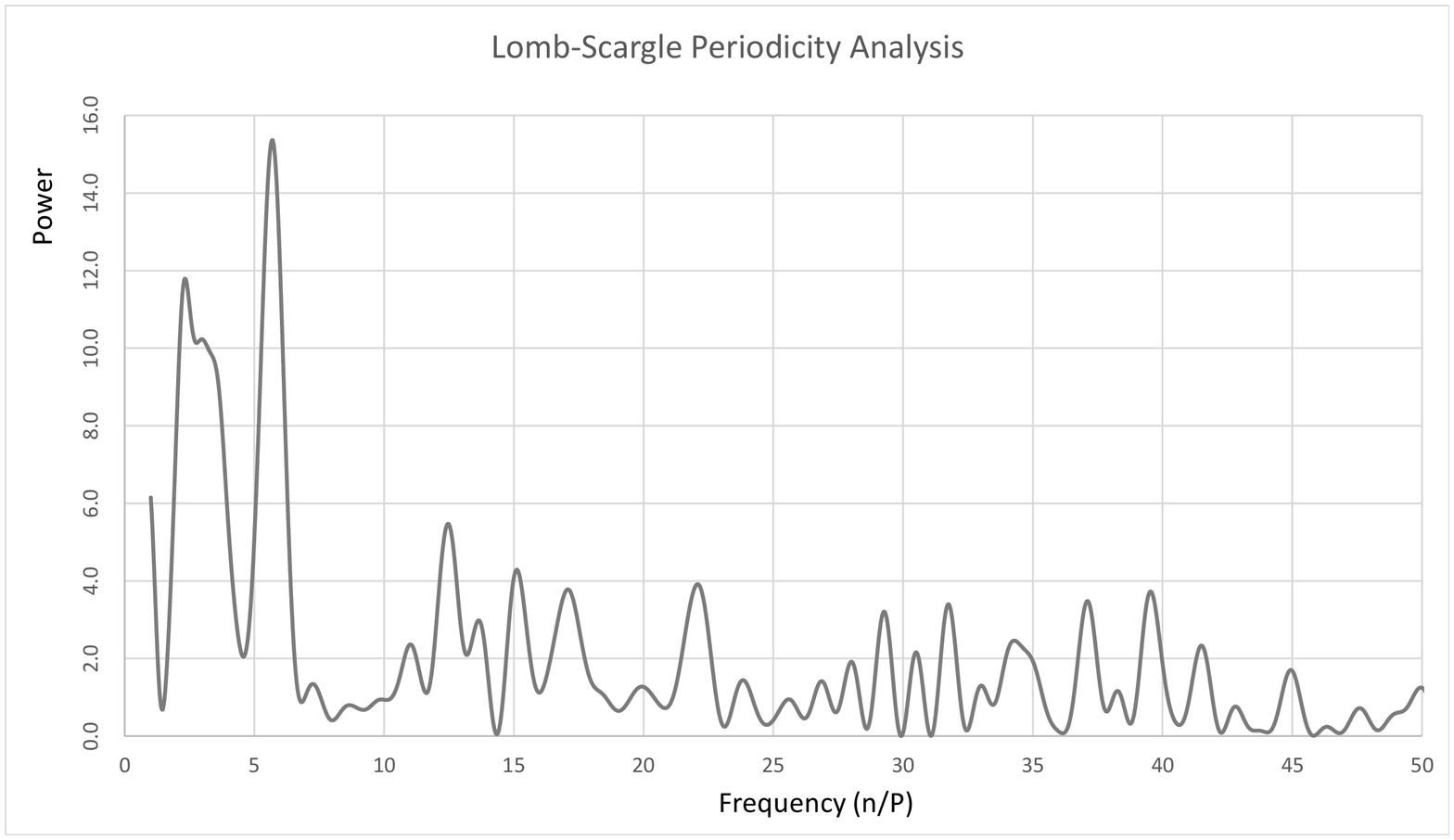} 
\label{fig:period}
\end{center}
\caption{Lomb-Scargle analysis of the power in the Kepler-91b light curve residuals 
against orbital frequency. Residuals here come from the differences between
the eclipsing binary model fitting to the whole, or full-phase, data-set; originally of $\sim$65000 points, but binned for this analysis to
a representative 360 points.  The high peak
at 6/$P$ is very distinct.}
\end{figure}

A periodicity analysis  was then carried out on the residuals from this fitting.  We show the result in Figure 3 (see also the residuals in Figs
4 and 5).
A very significant peak at $\sim$6/$P$ frequency can be seen, with a lower but wider peak at the 3/$P$ submultiple.
   The corresponding photometric contribution,  persisting
throughout the $\sim$240 orbits of the complete data-set, points to an orbit-oscillation resonance.  This may well correspond to 
the beat which is produced at $\sim$11.0 $\mu$Hz between the
higher frequency L2 and L0 modes shown by 
Lillo-Box et al (2014a), particularly
 the strong ones (n9L02 and n11L00)
 between $\sim$ 104 and 115 $\mu$Hz.
The existence of this additional  photometric effect is, of course, of interest in its own right;
its continuation evident in the orbitally phased and binned light curve even to the eye.  
The nature of this resonance
invites further physical discussion that we raise in Section 3.  
From the more immediate aim of parametrization 
of the close binary (star + planet) configuration, however, this is a complication that it
would be desirable to separate out.  Its retention in the analysis of 
Lillo-Box et al (2014a) 
may compromise the deduction of an eccentricity effect, or the parametrization in general,
to some extent.  

We therefore carried out a cleaning operation
by fitting the residuals from the original light curve fit with a Fourier decomposition reaching
to terms in frequency 6/$P$.  This  fitting is shown in Figure 4.  Our procedure is similar to that
used in the cleaning of close binary systems showing maculation effects (Zeilik et al., 1988).
In effect, we treat the light-curve oscillatory disturbance as a low-frequency superposition.  The higher frequency components that may also be present become strong at $\sim$10 times our frequency limit, but such high frequency components  are more akin to noise in their effects
on the eclipsing binary model.  Our final cleaned model and its fitting are shown in Figure 5.
We may note that the cleaned light curve shows a reduced eccentricity: a direct fitting of the raw data 
may thus be be open to doubt if 
interpreted simply as a real elliptical orbit.  At least, its determination in
a data-set containing both eclipsing binary and other causes of photometric variability
cannot be taken at face value.
%\begin{flushright}
%\end{flushright}

\begin{figure}[H]
\begin{center}
\includegraphics[width=\columnwidth]{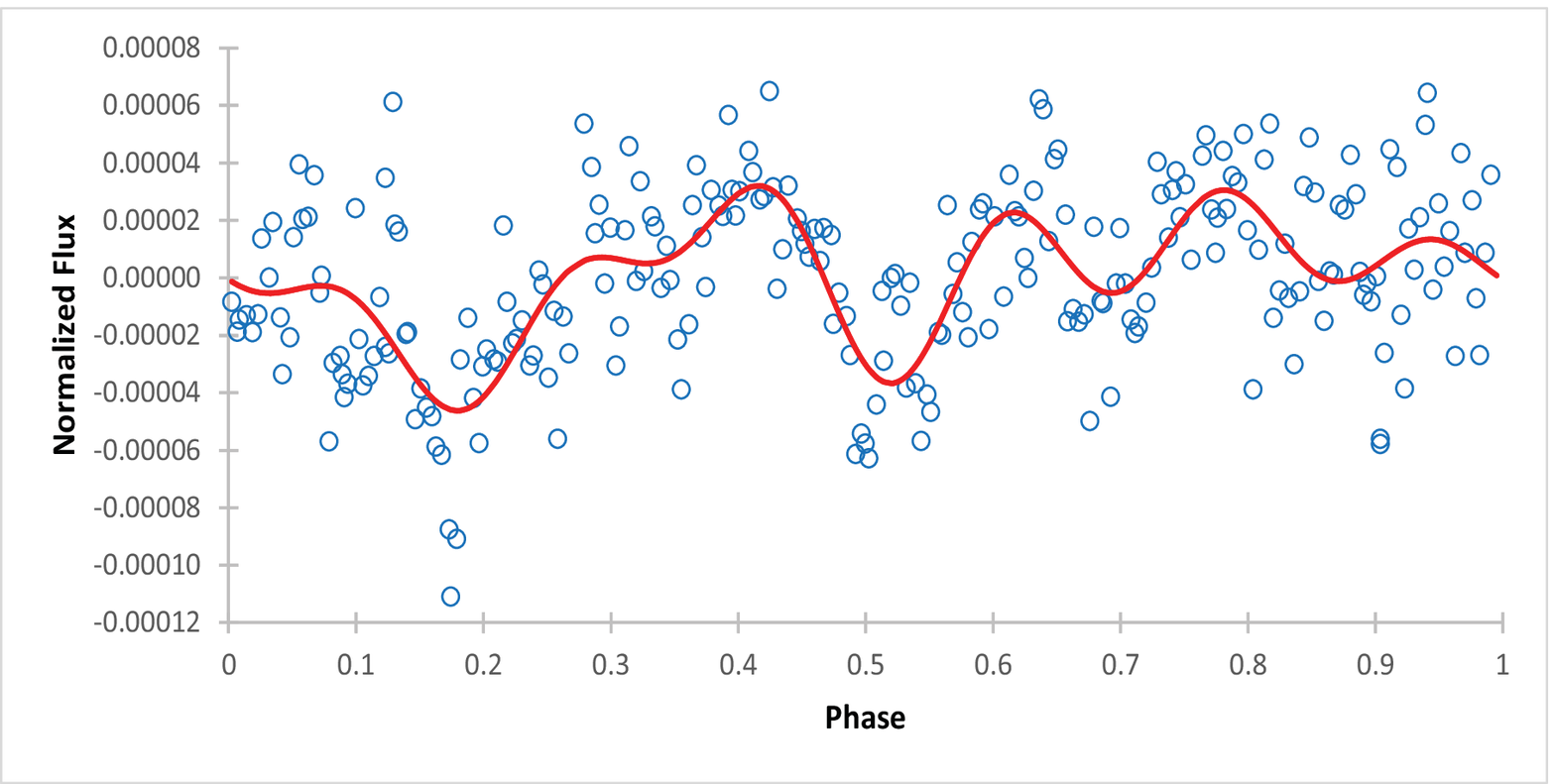}
\label{fig:cleanlc}
\end{center}
\caption{Fit of a Fourier series with terms up 
to 6 $\times$ the orbital frequency.  The dims at phases $\sim$1/6 and $\sim$1/2
look distinct and may suggest Trojan or Hilda type phenomena. This model
was added back with a negative sign to the original data-set. }  
\end{figure}
 
\begin{table*}
\begin{center}
\caption{Results of optimal fittings to various light curve data-sets.
\label{tbl-3}} 
{\footnotesize 
\begin{tabular}{lrrrrrrrrrr} 
 \multicolumn{1}{l}{ } & \multicolumn{1}{c}{$\Delta \phi_0$} 
 & \multicolumn{1}{c}{$r_1$~E1} & \multicolumn{1}{c}{$k$~E2} 
 & \multicolumn{1}{c}{$r_2$~E3} &	\multicolumn{1}{c}{$i$(deg)}&	
 \multicolumn{1}{c}{$q$~E4} & \multicolumn{1}{c}{$e$} & \multicolumn{1}{c}{$\omega$}  & \multicolumn{1}{c}{$\chi^2/\nu$}
& \multicolumn{1}{c}{$\Delta l$~E5}  \\ 
 \hline  	
TO   &	   --0.15(4)   & 4.08(7) & 2.21(3) & 8.9(2) &	 68.9(5)  &     	&   	&        &   & \\	  
BF   &	   --0.2(3)    & 4.03(6) & 2.16(2) & 8.7(2) &	 69.6(5)  &  4.2(8) &     	&	     &  1.19 & 2.5 \\	 	
BFe  &	   --0.3(2)    & 4.09(4) & 2.22(2) & 9.04(8)&	 69.8(4)  &  4.4(6) &0.05(2)& 300(15) & 1.07  &2.5 \\ 
BTO  &	   --0.1(1)    & 4.1 (2) & 2.2(1) & 8.9(4) &	 68.7(8)  &  7(7)   &	    &	      & 1.01  &2.2 \\	
BTOe &	   --0.1       & 4.1     & 2.1    & 8.8    &	 69       &  6      & 0.03  & 260    & 1.01  &2.2	 \\	  
%Mean &	   --0.15(1)   & 4.10(5)& 2.17(2) & 8.9(1) &	 69.1(5)  &  4.5(2) &	    & 	     &   &  \\	
NEAK &	              & 3.8(2)    & 2.159(+8,--31) & 8.2    &	 71.05	  &	        &       &        &   &	\\	
NEAC &	 		      &	         &	       &	         &	 67.4(6) &  4.8(6) &       &        &   &   \\	 
LB   &	              & 4.08(+1,-2)  & 2.26(+3.--10)&	         &	 69(1)&  6.4(+1,--2)	    &       &	     &  &\\	
     &                &           &          &           &        &         &       &        &  & \\	
\hline
\end{tabular}
}
\end{center} 

{\footnotesize			
\noindent Table 3 notation
																			
\noindent TO = Transit Only: means from the individual runs of each quarter	

\noindent BF = Binned Full: fitting to a single binning down to	360 points	

\noindent BFe = Binned Full: as above, but with eccentricity allowed		

\noindent BTO = Binned Transit Only: fitting of all transits, binned to 256 points	

\noindent BTOe = Binned Transit Only: as above, but with eccentricity allowed	
																											
\noindent NEAK = NASA Exoplanet Archive KOI Cumulative list		
	
\noindent NEAC = NASA Exoplanet Archive - Confirmed Planets

\noindent LB = Lillo-Box et al., (2013)	

Formal error estimates, where available, are 
indicated by the digits in parentheses. For 
reasons of space and convenient comparison in one panel, 
these are here truncated to the
rightmost digits in the parameter values.  
E$n$ means that the listed value
is 10$^n$ times the actual value. 
The unit of light values ($U$) for these fittings (not given here) are close to unity, with the very low scatter consistent with the errors 
indicated in Table 3 (i.e. $\sim$23 ppm for $\Delta l$ or $\sim$1 ppm for $\Delta U$).

The BFe row of this table is adopted for our final presentation in Table 4.		
We present the reduced $\chi^2/\nu$ value for this and the BF solution with 
the same error estimate of 0.000025 for the normalized 360 point binned data-set.  The decrease of  $\chi^2/\nu$
from 1.19 to 1.07 is significant at the 95\% confidence level, i.e.\ there is a relatively high probability to
to the improvement produced by an eccentric orbit model.   This is not the case
for the transit-only fittings, where, though modelling is
 effective in reducing the relative scatter of the residuals, the parameters are
 less well defined.  In fact, if eccentricity is allowed to be simultaneously adjustable
 the $\chi^2$ Hessian becomes non-positive-definite, so that regular error estimates lose meaning.
}
\end{table*}

\begin{figure}[H]
\begin{center}
\includegraphics[width=\columnwidth]{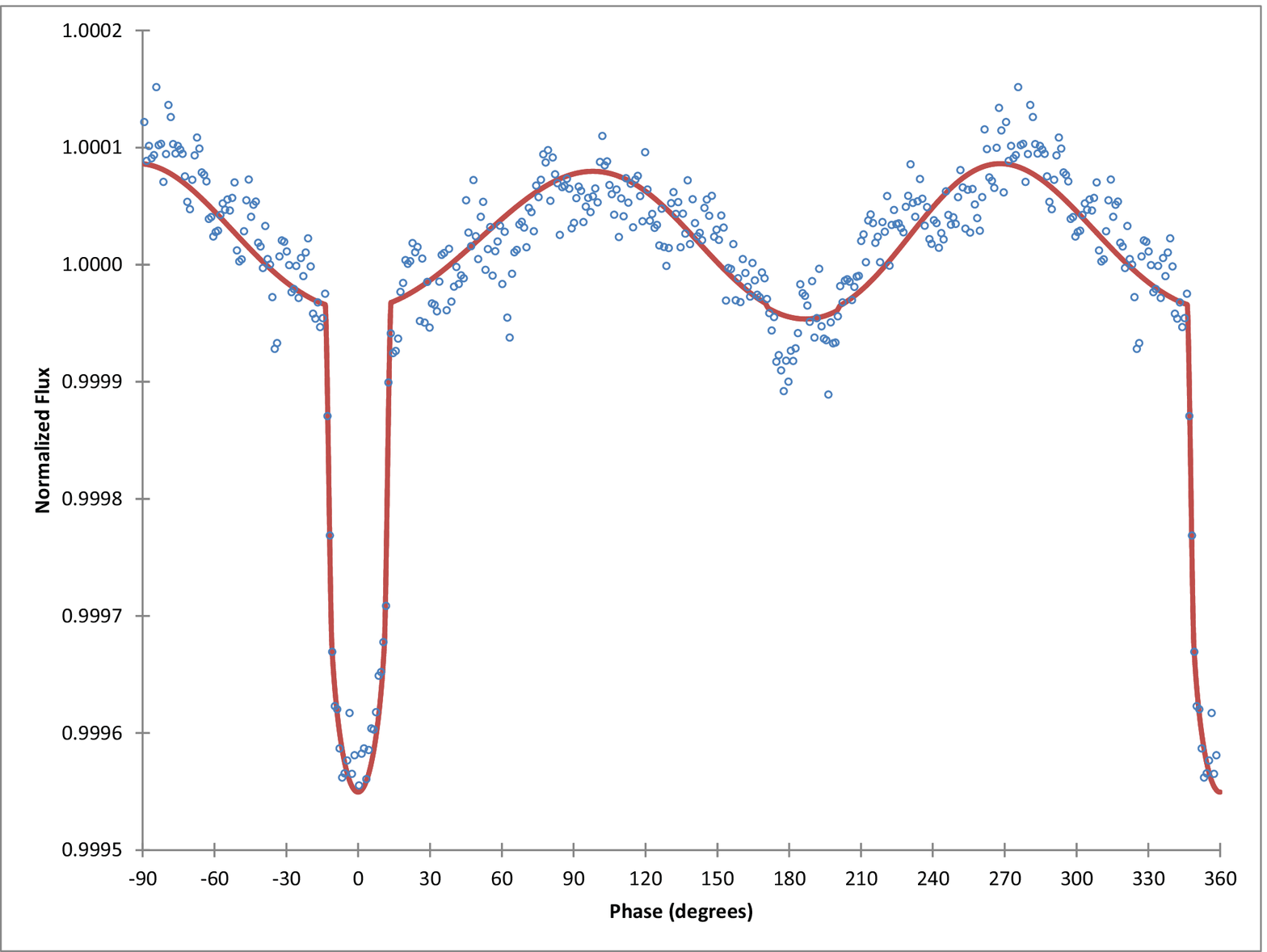}
\includegraphics[width=\columnwidth]{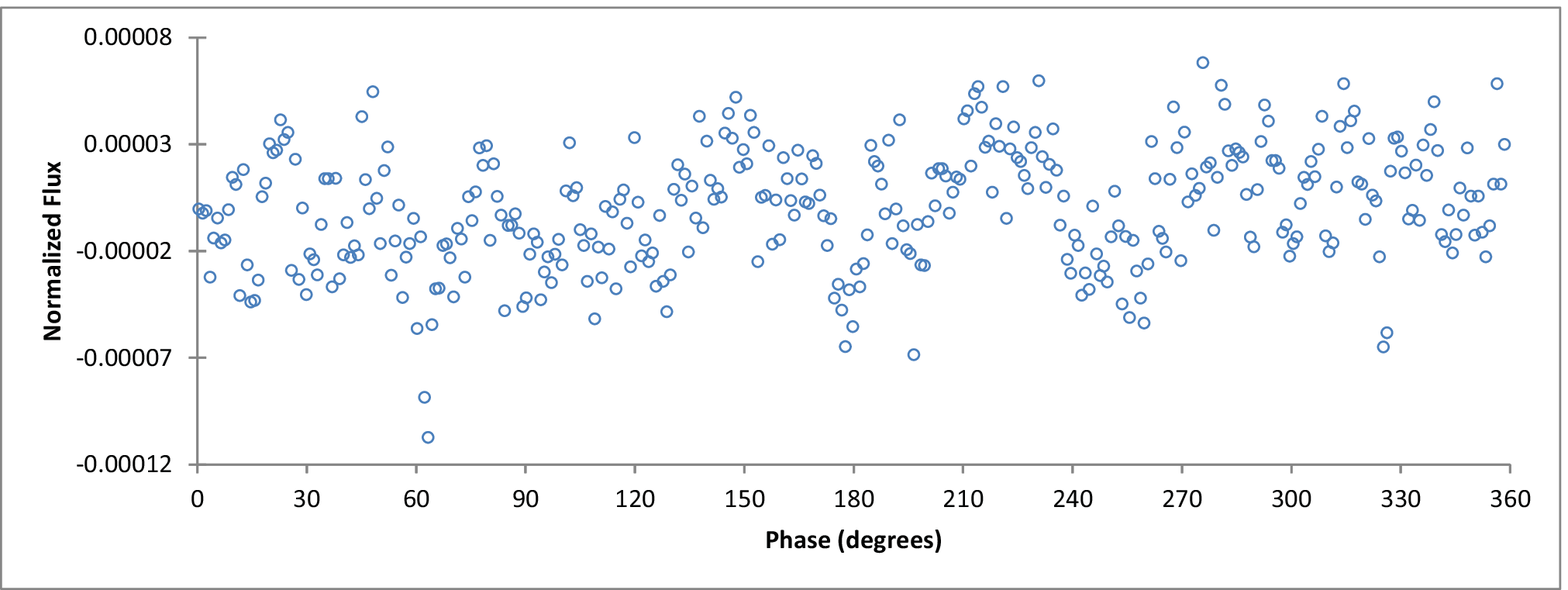}
\label{fig:Kep91binlc}
\end{center}
\caption{Complete cleaned and binned light curve of Kepler-91, showing proximity effects and an
apparent secondary minimum, together with the {\sc WinFitter} model.
The residuals are shown below.
This may be compared with Fig 7 in Lillo-Box et al.\ (2014a). } 
\end{figure}

 In Table 4 our final results and their implications for absolute parameters are listed and compared with the results of other authors.

\begin{table*}
\begin{center}
\caption{We list here the Lillo-Box et al.\ (2014a) (LB) parameters, as well as those of Barclay et al.\ (2015), Sato et al.\ (2015) and Esteves et al.\ (2015),  alongside
those of the present paper (P). Error estimates, influenced  by the compilation of results, are listed in the final column. 
Where no error is listed the corresponding parameter has been taken from
information separate to the fitting.
% [{\bf \color{red}  My Excel file KOI 2331 parameters comparison.xlsx shows the values of the parameters for each of these. }]
\label{tbl-3}} 
 \begin{tabular}{lcccccc}
\hline
\multicolumn{1}{c}{Parameter} & \multicolumn{1}{c}{LB} & \multicolumn{1}{c}{Barclay} & \multicolumn{1}{c}{Sato} &
\multicolumn{1}{c}{Esteves} &
\multicolumn{1}{c}{P} & \multicolumn{1}{c}{Error}\\
\hline \
Epoch & 2454969.3966 & 2454969.3837 & 2454969.3866 & 2454969.3958 & --- & \\
P  (d)      & 6.24658 &    &   6.24668005 & 6.24658 & ---  & \\   
dist.\ (pc) & 1030 & & & & --- 
& \\
$M_s$ ($\odot$) & 1.31 & 1.31 & 1.31 & 1.31 &1.3 & \\
$T_p$ (K) & 2000 & & 1920 - 2460 & 2040 - 3000 & 2000 &\\
$a$ $\odot$ & 15.62 & 15.53 & 15.93& 15.71 & 15.57 & \\
log $g_s$ & 2.95 & &  & 2.953& 2.96 &\\
$M_p/M_s$ & 0.00061 & 0.00053 & 0.00048 & 0.00059& 0.00044 & 0.00006\\
$r_s$ (mean)& 0.403 & 0.406 & 0.406 &0.401& 0.409 & 0.004\\
$r_p$ (mean)& 0.0091 & 0.0088 & 0.0090 & 0.0087 & 0.0090 & 0.0002 \\
$i$ (deg)& 67-78 & 69.17 & 67.37 & 69.7& 69.8 & 0.5 \\
$e$ & 0.07 & 0.02 & 0.04 & 0.0 & 0.05 & 0.02 \\
$\omega$ (deg) & 320.0 & 41.9 & 302.8 & & 300 & 20 \\
$b$ & 0.88 & 0.87 & 0.90 & 0.87&  0.85 & 0.03\\
\hline
\end{tabular}
\end{center}
\vspace{1ex}
{\footnotesize In the P column --- indicates values as in LB.} 
\end{table*}
 
\newpage
\section{Discussion}

\noindent {\bf 3.1 \hspace{2em} Mass ratio}

The preceding presentation on the {\em Kepler} Mission's photometric data on 
the Kepler-91 system to a large extent supports the picture given
by Lillo-Box et al.\ (2014a, b), Barclay et al.\ (2015), Sato et al.\ (2015) and Esteves et al.\ (2015).
 The close binary system photometry analysis program  {\sc WinFitter} that
 we have adapted and applied, however, includes a fuller description
  of the surface distortions due to tides and rotation, as well
  as the reflection effect, than has generally been used in similar studies hitherto.
 Given the relatively large fractional size of the star in the present case
 the higher power terms are significant and we can expect
  a systematic improvement in the mass ratio (reduced by $\sim$ 40\% from
 that determined by Lillo-Box et al., 2014a). Our derived value $q \approx 0.00044 \pm 0.00006$ for the
 mass ratio is then in good agreement with that which can be 
 inferred from the relatively high-accuracy radial velocity data of Barclay et al.\ (2015)
 and perhaps even more with the greater coverage and higher mean accuracy results 
 of Sato et al.\ (2015).
 
 \vspace{2ex}
 \noindent {\bf 3.2 \hspace{2em} Additional effects}
 
It is clear from the comparison of the observed scatter in the data compared with
the photon noise expectable for  a $V \approx 12.88$ star that there are additional
short term variations, which can be associated with the non-radial
pulsational spectrum also studied by Lillo-Box et al (2014a). In fact, the scatter in
the raw data makes individual planetary transits only 1$\sigma$ events: binning by a factor of $\sim$100
therefore appears desirable to allow a reasonably confident approach to parametrization.
From the empirical point of view, the light curve produced in this way can be characterized
by, in addition to regular eclipsing binary system effects, 
(a) an `O'Connell effect' (asymmetric maxima) and (b) the persistence of quasi-periodic
small light drops or dims. 
The former can be associated with the effect of orbital eccentricity,
given the slight, though calculable, effects of a correspondingly varying tidal distortion of the star.
The possibility of photometric effects associated with some surface maculation may 
be mentioned, given the cool temperature and likely structure of this star.
We know, however, from the slow rotation speed derived by Lillo-Box et al (2014a) that
the planet performs around 7.6 revolutions in one rotation of the star
(unlike the synchronized arrangement usually found in close binary stars
-- cf.\ e.g.\ Zahn, 1977).
Maculation seems thus unlikely to persist as coherent systematic effects
in the binned light curve in general, though starspots may be partly associated with the
relatively high scatter.  

\vspace{2ex}
\noindent {\bf 3.3 \hspace{2em} Explaining the dims }

Regarding the dims, Lillo-Box et al.\ raised possible lines of explanation involving
  other planets or satellites that 
co-rotate synchronously or are in a resonant arrangement with Kepler-91b.
The idea of eclipses of the planet's reflected light, although apparently 
modelled by Barclay et al.\ (2015), 
 faces at least conceptual difficulties,
since the dims are more than twice as deep as the regular scale of 
reflected light at full phase. 
When this  is eclipsed out at the secondary minimum the light level should not be appreciably lower than at phases just outside the primary
minimum; the star's light only being received in either case.
We have to countenance also improbability going with the paired circumstances of another planet's
non-coplanar orbit, though still near the line of sight, which is {\em also} at a
resonant period.  

Given the v-shaped effect at phase $\sim$60\degr
the possibility of a large Trojan concentration can be considered: similarly, perhaps a Hilda concentration could be associated with the extra light loss
about the secondary minimum.  Such situations have been examined, for example by Schwarz et al.\ (2007), or in the volume edited by
Souchay \& Dvorak (2010), but more detailed clarifying explanation is needed.  It seems clear that additional systematic effects are present in the light curve that survive the repeated foldings and should therefore have some connection to the orbit.  However, it is also seen that the depth of the dims is $\sim$1/10 that of the transit, entailing that 
any planetoid concentration must involve the equivalent of at least
several thousand Vesta-sized objects if it is to fully account for the scale of light loss.  The effect is several orders of magnitude greater than could apply to the Trojan concentration in the Solar System 
(Yoshida \& Nakamura, 2005), and so must be regarded cautiously.

Lillo-Box et al.\ (2014a) mentioned subtle effects  
associated with the {\em Kepler} data-processing.  
A known effect associated with the relatively
large pixel size of the {\em Kepler} detectors is the presence of a remote
Algol system in the field.   There would again be a paired coincidence
required in such a case associated with the orbital period integral sub-multiplicity.  Additionally, false-positive scenarios of this kind 
diminish to low probability following high-resolution analysis 
of data from the AstraLux
lucky-imaging camera on the 2.2 m telescope at the Calar Alto Observatory
included in the study of Lillo-Box et al.\ (2014a).
The more direct line of interpretation  raised in the previous section
is thus favoured, though the suggested pipeline error possibility cannot be entirely ruled out.

 Our treatment, introducing a low pass filter to remove these apparently systematic and orbitally resonant
 effects, leads to more precise eclipsing binary model defined parameters,
 as with the approach of Barclay et al.\ (2015), as well as casting doubt on the precise representation
 of light curve asymmetry in terms of simply an elliptical orbit.
 Nevertheless, the light curve asymmetry is maintained even after cleaning
and the most direct explanation of that is in terms of a slightly eccentric orbit,
even given an expectable trend towards orbit circularization (Zahn, 1977),
keeping in mind the great age of the system estimated by Lillo-Box et al.\ (2014a).
At this point, we may note the possibility of orbital eccentricity -- resonant pulsational
interactions, as considered, for example, in \
heartbeat' pulsators (as an extreme case: see e.g.\ Hambleton et al., 2013).
Physically, the argument is that an orbital eccentricity that might otherwise be frictionally
eroded can be trapped at a certain value if appropriate amounts of energy are fed back into the orbit
at just the right times
from a suitable $\kappa$-mechanism  (Henrard, 1982; Alexander, 1988; Witte and Savonije, 1997). However, the
potentially resonant process here is actually a beat effect, rather than a direct eigenfrequency of the underlying kappa-mechanism.  Even so, the amplitude envelope of selected strong eigenfunction pairs does apparently resonate at 6/P with the orbital frequency.  
 That the suggested mechanism is, or is not, possible physically, thus invites
fuller follow-up theoretical consideration.

\vspace{2ex}
\noindent {\bf 3.4 \hspace{2em} Period variation}

 Our modelling suggested the possibility of a steady decline in the period, although the
 likelihood of this is not high, given the uncertainties of the time of minimum determination.
On the other hand, there appears some suggestion, from the results of Sato et al.\ (2015) that
the period may be increasing in the short term.  In a similar way, an apparent trend to decrease of the
longitude of periastron over the full range of the {\em Kepler} Mission surveillance
cannot be taken so seriously given the relatively high errors of the determination. 
 
But, in connection with such effects, Budding (1984) gave a possible form for the variation of the separation $A$
between the components of a binary system with total mass $M$ and two stellar masses $m_1$, $m_2$, in which mass loss and transfer
was taking place as 
\begin{equation}
\frac{\dot{A}}{A} = 2 \left\{ \frac{\dot{J}}{J} - \frac{\dot{m_1}}{m_1} - \frac{\dot{m_2}}{m_2}
- M(\dot{\epsilon_1}  +   \dot{\epsilon_2} + \dot{\epsilon_3})  \,\,\,   ,
 \right\}
\end{equation}
where $J$ is the system's angular momentum, and the small components in $\epsilon_i$ are
associated with the rotational changes of the two bodies and some surrounding matter.
These latter components are often neglected, at least in a preliminary assessment.

In classical Algols, regarded as involved in a process of {\em mass transfer} and where $m_1$
denotes the originally more massive and mass-losing star,
this equation can be simplified to 
\begin{equation}
\frac{\dot{A}}{A} \approx 2 \left\{ \frac{\dot{J}}{J} - \frac{\dot{m_1}}{m_1}.\frac{(M - 2m_1)}{(M-m_1)}\right\}
\end{equation}
where we have written simply $\dot{m_2} = -\dot{m_1}$. In a mass loss process, $\dot{m_1}$ is, by definition negative, so this equation tells us that once $m_1$ has dropped below half the
system's mass, further mass loss will widen out the separation, unless there is some 
significant loss of system angular momentum through $\dot{J}/J$.  This is usually considered
small for classical Algols in the process of conservative mass transfer, so the separation
of the components in this phase is expected to increase.
If $m_1$ were to become very small, but still engaging in mass transfer in this way Eqn 2
has a small denominator, which would tend to amplify the separative effect of even a low rate of mass transfer.  
It is feasible that the planet Kepler-91b might be losing mass in this way
through the atmospheric inflation process considered by Lillo-Box et al.\ (2014a).
The mean density of the planet certainly seems low in comparison to more typical
hot jupiters.  This might then be consistent with the apparent period increase indicated
by Sato et al (2015).

Various possibilities that might produce a noticeable variation of $\dot{J}/J$ could be considered,
including tidal friction and magnetic braking.  The latter 
offers a wide range of possible parameter variation and appears able to introduce significant effects
in a variety of situations.  In general, of course, we should expect such frictional effects to 
make for a negative $\dot{J}/J$, which would entail period decrease, as suggested in Figure 2.
Reasonable estimates for $\dot{J}/J$  due to magnetic braking were estimated by Demircan et al.\ (2006)
and also Erdem \& \"{O}zt\"{u}rk (2014).  However, the scale of effect that might be interpreted
from Figure 2 has $\dot{P}/P \sim 10^{-7}$, which is a 2-3 orders of magnitude greater
than the effects considered in these papers.  
Granted that the scale of error in Fig.\ 2 makes any present attempt at interpretation
in physical terms unrealistic, still the {\em possibility} of period changes is intriguing.
A magnetically driven mass loss, tending to decrease the orbital period, could be reasonably
expected from this star; while mass transfer from the inflated planet
 would increase the period.
The combination of effects in the Kepler-91 system overall thus
 make for interesting alternatives, and the case for continued surveillance
 is compelling.

\acknowledgements This research has been supported by TUBITAK (Scientific and Technological Research Council of Turkey) under Grant No.\ 113F353.  The help of colleagues in the Physics Dept., COMU, including Dr 
M.\ T\"{u}ys\"{u}z, is acknowledged.

\end{document}